\documentclass{ws-ijmpcs}

\usepackage{amsmath}
\usepackage{latexsym}
\usepackage{amssymb}
\usepackage{dsfont}

\newcommand{\ud}{\mathrm{d}}

\begin{document}

\markboth{C. Lorc\'e and B. Pasquini}
{Wigner distributions and quark OAM}

%
\catchline{}{}{}{}{}
%

\title{WIGNER DISTRIBUTIONS AND QUARK ORBITAL ANGULAR MOMENTUM}

\author{C\'EDRIC LORC\'E}

\address{IPNO, Universit\'e Paris-Sud, CNRS/IN2P3, 91406 Orsay, France\\
and LPT, Universit\'e Paris-Sud, CNRS, 91406 Orsay, France\\
lorce@ipno.in2p3.fr}

\author{BARBARA PASQUINI}

\address{Dipartimento di Fisica, Universit\`a degli Studi di Pavia, Pavia, Italy\\
and Istituto Nazionale di Fisica Nucleare, Sezione di Pavia, Pavia, Italy\\
Barbara.Pasquini@pv.infn.it}

\maketitle

\begin{history}
\received{Day Month Year}
\revised{Day Month Year}
\end{history}

\begin{abstract}
We discuss the quark phase-space or Wigner distributions of the nucleon which combine in a single picture all the information contained in the generalized parton distributions and the transverse-momentum dependent parton distributions. In particular, we present results for the distribution of unpolarized quarks in a longitudinally polarized nucleon obtained in a light-front constituent quark model. We show how the quark orbital angular momentum can be extracted from the Wigner distributions and compare it with alternative definitions.
\end{abstract}


\section{Introduction}	

Unraveling the internal structure of hadrons is a challenging but crucial problem as it provides important information about the non-perturbative regime of quantum chromodynamics (QCD). For certain high-energy processes, the description can be factorized into a process-dependent but perturbative part and a process-independent but non-perturbative part. The latter contains the information we are looking for, namely how the partons are distributed in the nucleon and how they are correlated.

The quantum phase-space or Wigner distributions encode in a unified picture the information obtained from the transverse-momentum dependent parton distributions (TMDs) and the generalized parton distributions (GPDs) in impact-parameter space. The concept of Wigner distributions in QCD was first explored in Refs.~\refcite{Ji:2003ak} and \refcite{Belitsky:2003nz} where relativistic effects were neglected. Recently, we identified the impact-parameter representation of the generalized transverse-momentum dependent parton distributions\cite{Meissner:2009ww} (GTMDs) with the five-dimensional Wigner distributions (two position and three momentum coordinates) which are not plagued by relativistic corrections\cite{Lorce:2011kd}. Even though the Wigner distributions do not have a strict probablisitic interpretation due to the uncertainty principle, they encode a variety of information and can often be interpreted with semiclassical pictures.

In this contribution, we discuss the phenomenology of the quark Wigner distributions based on successful relativistic quark models\cite{Lorce:2011dv}, since so far it is not known how to access these distributions directly from experiments. We focus in particular on how the quark orbital angular momentum (OAM) can be extracted from the Wigner distributions, and compare it with alternative definitions based on the GPDs and the TMDs.

\section{Parton correlation functions}

The maximum amount of information on the quark distribution inside the nucleon is contained in the fully-unintegrated quark-quark correlator $\tilde W$ for a spin-$1/2$ hadron\cite{Ji:2003ak,Belitsky:2003nz,Meissner:2009ww}, defined as
\begin{equation}\label{gencorr}
\tilde W^{[\Gamma]q}_{\Lambda'\Lambda}(P,k,\Delta,n)=\frac{1}{2}\int\frac{\ud^4z}{(2\pi)^4}\,e^{ik\cdot z}\,\langle p',\Lambda'|\overline\psi(-\tfrac{z}{2})\Gamma\,\mathcal W\,\psi(\tfrac{z}{2})|p,\Lambda\rangle.
\end{equation}
This correlator is a function of the initial and final hadron light-front helicities $\Lambda$ and $\Lambda'$, respectively, the average hadron and quark four-momenta $P=(p'+p)/2$ and $k$, respectively, and the four-momentum transfer to the hadron $\Delta=p'-p$. The superscript $\Gamma$ stands for any element of the basis $\{\mathds 1,\gamma_5,\gamma^\mu,\gamma^\mu\gamma_5,i\sigma^{\mu\nu}\gamma_5\}$ in Dirac space. A Wilson line $\mathcal W\equiv\mathcal W(-\tfrac{z}{2},\tfrac{z}{2}|n)$ ensures the color gauge invariance of the correlator, connecting the points $-\tfrac{z}{2}$ and $\tfrac{z}{2}$ \emph{via} the intermediary points $-\tfrac{z}{2}+\infty\cdot n$ and $\tfrac{z}{2}+\infty\cdot n$ by straight lines, where $n$ is a light-like four-vector. This naturally induces a dependence of the Wilson line on the light-front direction $n$.

Since the parton light-front energy $k^-$ is particularly difficult to access in high-energy experiments, the relevant correlators are actually integrated over $k^-$, setting all the fields at the same light-front time $z^+=0$
\begin{align}\label{GTMDcorr}
W^{[\Gamma]q}_{\Lambda'\Lambda}&(P,x,\vec k_\perp,\Delta,n)=\int\ud k^-\,\tilde W^{[\Gamma]q}_{\Lambda'\Lambda}(P,k,\Delta,n)\nonumber\\
&=\frac{1}{2}\int\frac{\ud z^-\,\ud^2z_\perp}{(2\pi)^3}\,e^{ixP^+z^--i\vec k_\perp\cdot\vec z_\perp}\,\langle p',\Lambda'|\overline\psi(-\tfrac{z}{2})\Gamma\,\mathcal W\,\psi(\tfrac{z}{2})|p,\Lambda\rangle\big|_{z^+=0},
\end{align}
where we used for a generic four-vector $a^\mu=[a^+,a^-,\vec a_\perp]$ the light-front components $a^\pm=(a^0\pm a^3)/\sqrt{2}$ . Furthermore, in Eq.~\eqref{GTMDcorr} $x=k^+/P^+$ and $\vec k_\perp$ are the average fraction of longitudinal momentum and average transverse momentum of the quark, respectively. This correlator is parametrized in terms of the GTMDs\cite{Meissner:2009ww}. While they are very useful from a theoretical point of view, only particular sections and projections of these functions are known to be accessible experimentally. 

\begin{figure}[h]
	\centerline{\psfig{file=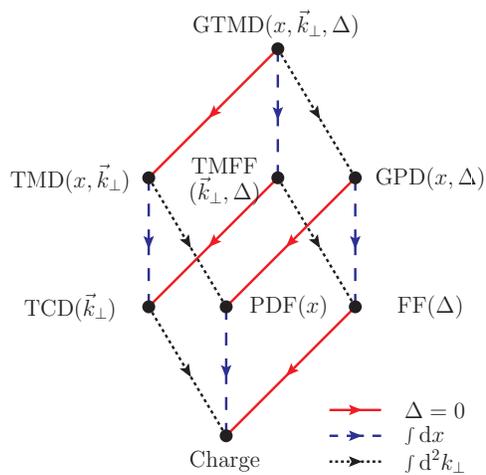,width=.5\textwidth}}
	\caption{Representation of the different sections and projections of the GTMDs. The solid, dashed and dotted arrows correspond to the forward limit $\Delta=0$, the integration over the quark longitudinal and transverse momentum, respectively.\label{figbox}}
\end{figure}

In semi-inclusive high-energy experiments, like \emph{e.g.} semi-inclusive deep inelastic scattering  and Drell-Yan processes, one can access the TMDs corresponding to the forward limit $\Delta=0$ of the GTMDs. In exclusive high-energy experiments, like \emph{e.g.} deeply virtual Compton scattering and deeply virtual meson production, one can access the GPDs corresponding to the first $\vec k_\perp$ moment of the GTMDs. The forward limit of the GPDs and the first $\vec k_\perp$ moment of the TMDs coincide, leading to the so-called parton distribution functions (PDFs), and can be accessed in deep inelastic scattering. To these GTMDs, TMDs, GPDs and PDFs, correspond $x$-integrated functions called  transverse-momentum dependent form factors (TMFFs), transverse charge densities (TCDs), form factors (FFs) and charges, respectively. A summary of the relations between the different distributions is shown in Fig.~\ref{figbox}. For more details, see Refs.~\refcite{Meissner:2009ww} and \refcite{Lorce:2011dv}.

\section{Wigner distributions}

The GTMDs have in general no clear partonic interpretation unless we consider them at the leading twist $\Gamma=\gamma^+,\gamma^+\gamma_5,i\sigma^{j+}\gamma_5$ with $j=1,2$ and perform a Fourier transform to the impact-parameter space\cite{Soper:1976jc,Burkardt:2000za,Burkardt:2002hr}. We then obtain quark distributions which are naturally interpreted as Wigner distributions\cite{Ji:2003ak,Belitsky:2003nz,Lorce:2011kd}
\begin{equation}\label{wigner}
\rho^{[\Gamma]q}_{\Lambda'\Lambda}(x,\vec k_\perp,\vec b_\perp,n)\equiv\int\frac{\ud^2\Delta_\perp}{(2\pi)^2}\,e^{-i\vec\Delta_\perp\cdot\vec b_\perp}\,W^{[\Gamma]q}_{\Lambda'\Lambda}(P,x,\vec k_\perp,\Delta,n).
\end{equation}
Although the GTMDs are in general complex-valued functions, their two-dimensional Fourier transforms are always real-valued functions, in accordance with their interpretation as phase-space distributions. We note that, like in the usual quantum-mechanical Wigner distributions, $\vec b_\perp$ and $\vec k_\perp$ are not Fourier conjugate variables. However, they are subjected to Heisenberg's uncertainty principle, because the corresponding quantum-mechanical operators do not commute $[\hat{\vec b}_\perp ,\hat{\vec  k}_\perp] = 0$. As a consequence, the Wigner functions can not have a strict probabilistic interpretation.

There are in total 16 Wigner functions at twist-two, corresponding to all the 16 possible configurations of nucleon and quark polarizations. For simplicity, we focus here on the longitudinal polarization, but other configurations for the quark and proton polarizations can be found in Ref.~\refcite{Lorce:2011kd}. The unpolarized quark distribution inside a longitudinally polarized nucleon is given by the Wigner distribution $\rho^{[\gamma^+]q}_{\Lambda\Lambda}(x,\vec k_\perp,\vec b_\perp,n)$. This distribution contains a nucleon spin-independent contribution $\rho^q_{UU}\equiv\rho^{[\gamma^+]q}_{\Lambda\Lambda}(x,\vec k_\perp,\vec b_\perp,n)+\rho^{[\gamma^+]q}_{-\Lambda-\Lambda}(x,\vec k_\perp,\vec b_\perp,n)$ and a nucleon spin-dependent contribution $\rho^q_{LU}\equiv\rho^{[\gamma^+]q}_{\Lambda\Lambda}(x,\vec k_\perp,\vec b_\perp,n)-\rho^{[\gamma^+]q}_{-\Lambda-\Lambda}(x,\vec k_\perp,\vec b_\perp,n)$. For obvious symmetry reasons, only the contribution $\rho^q_{LU}$ contains the information about the quark OAM. The quark longitudinal OAM originates from the correlation between the quark transverse position and momentum, and is therefore not accessible at leading twist \emph{via} the GPDs or the TMDs.

\begin{figure}[h!]
	\centerline{\psfig{file=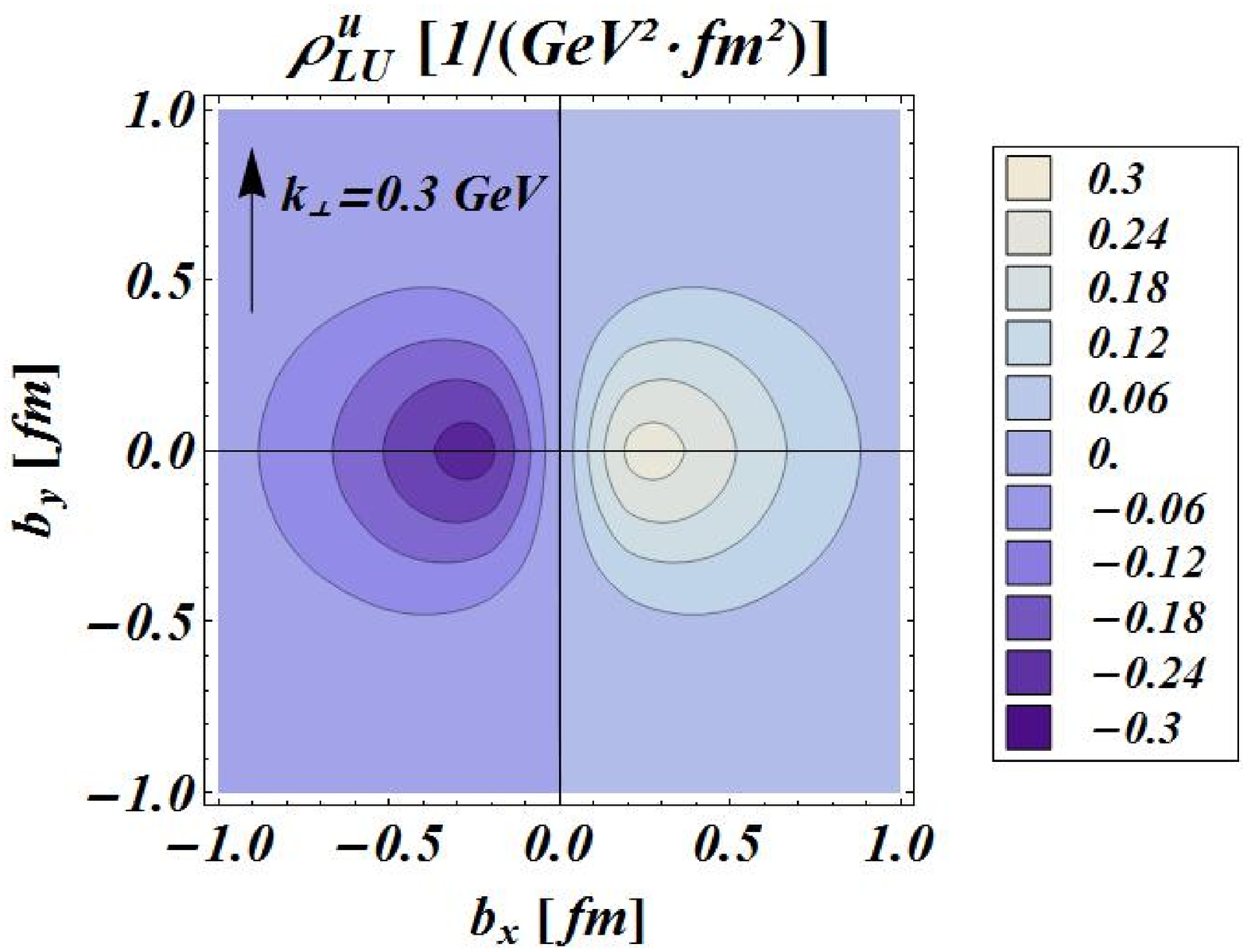,width=.5\textwidth}	\psfig{file=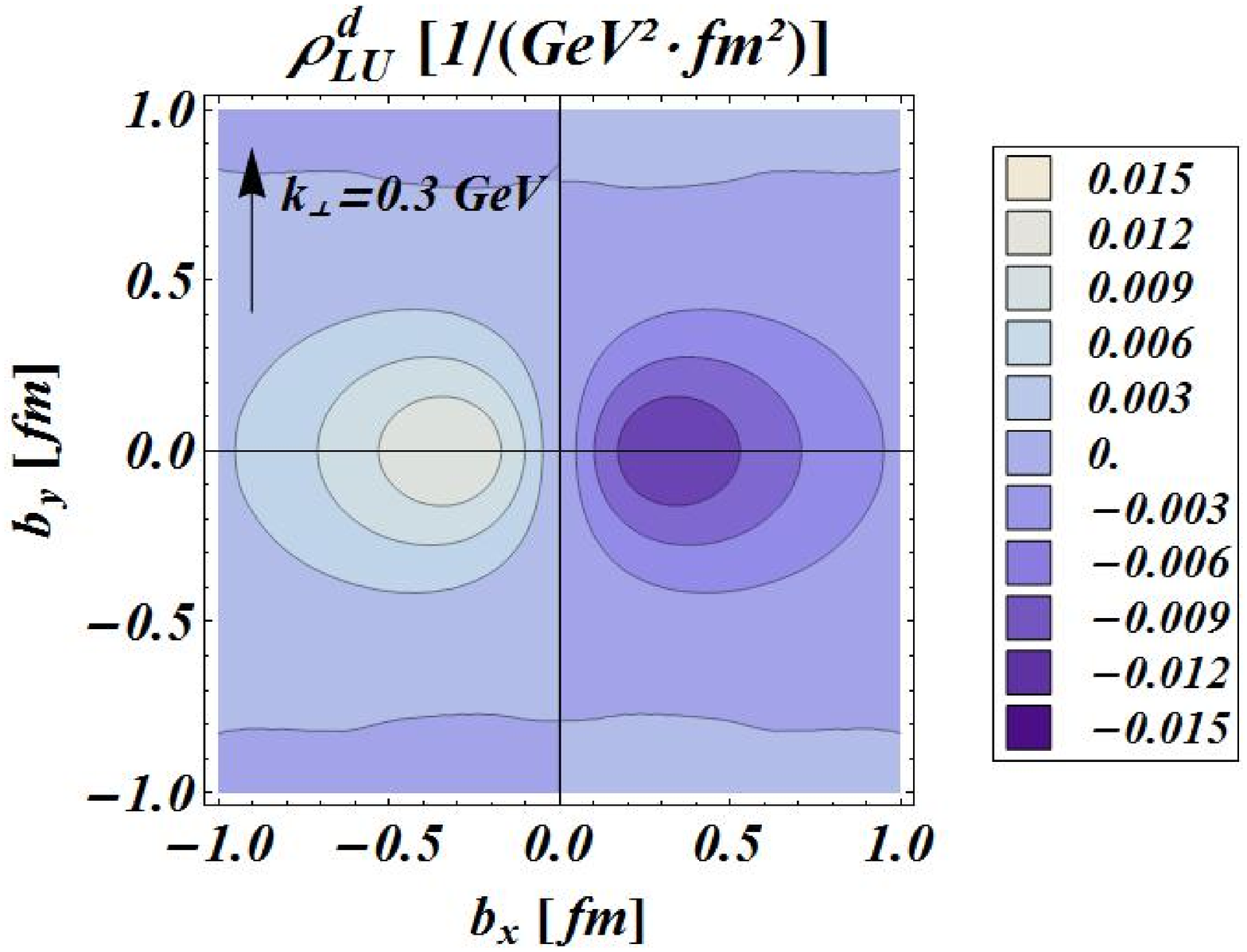,width=.5\textwidth}}
           \centerline{\psfig{file=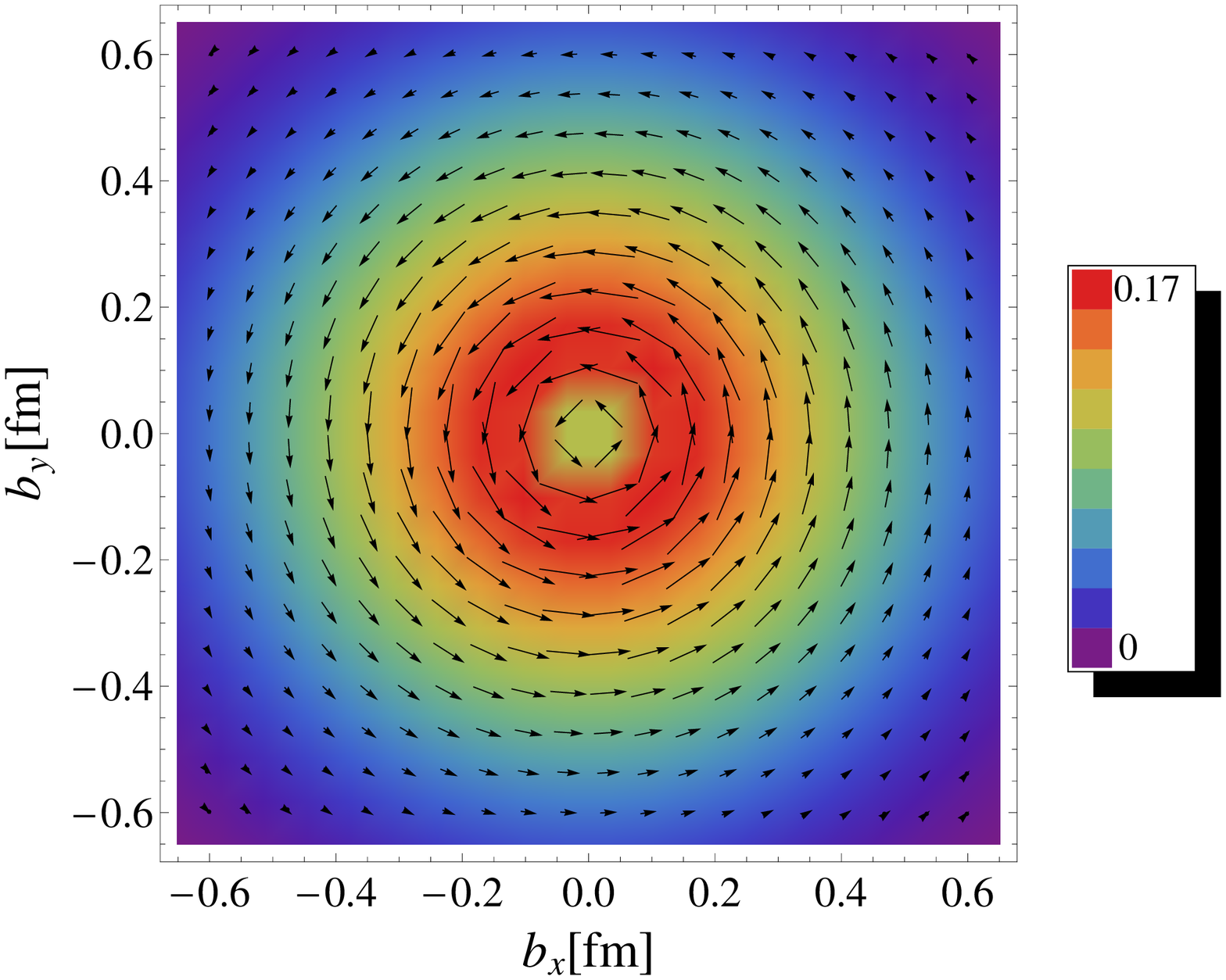,width=.42\textwidth}\hspace{1.2cm}\psfig{file=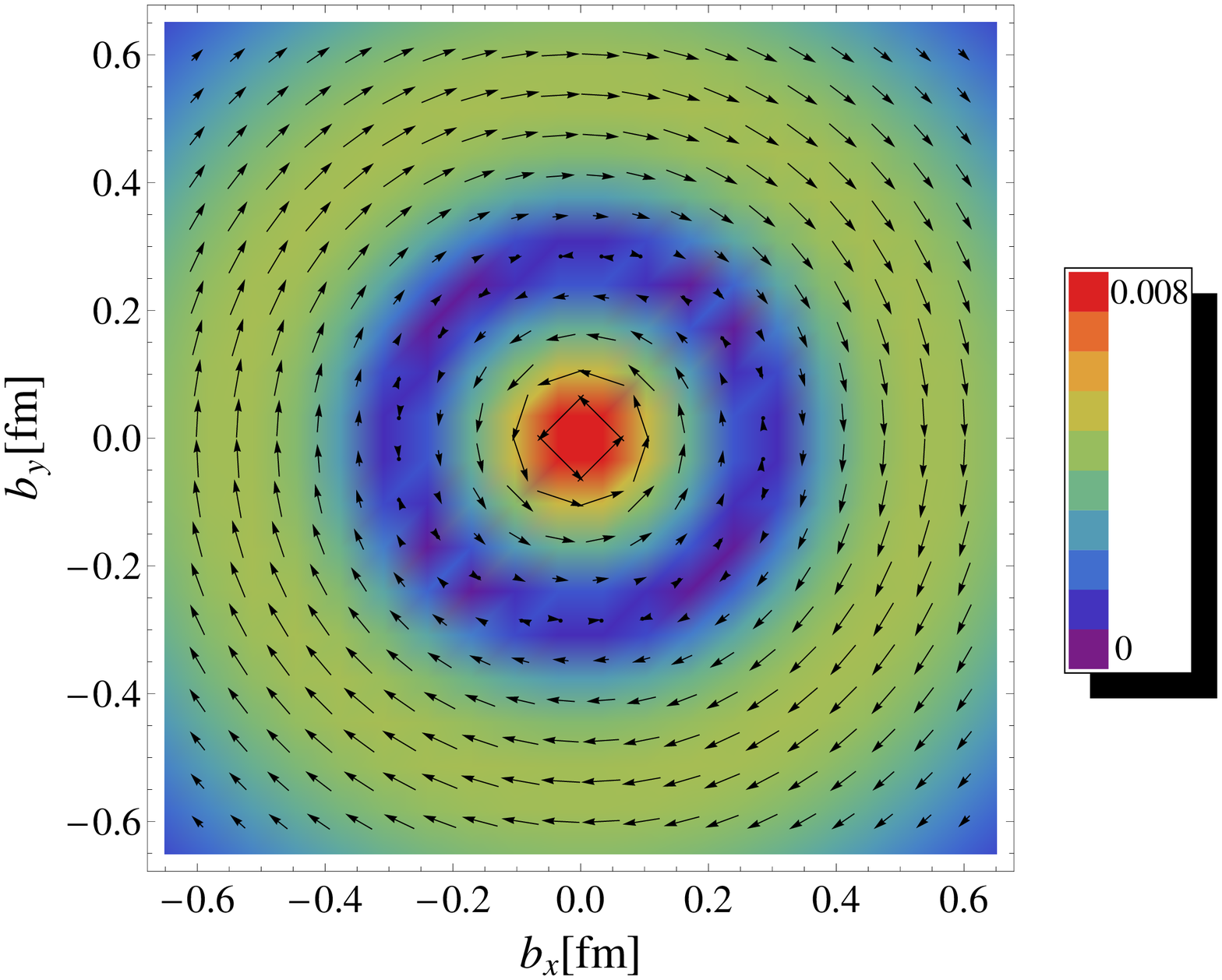,width=.42\textwidth}}
	\caption{The $x$-integrated distributions in impact-parameter space for unpolarized quarks in a longitudinally polarized proton (the proton spin points out of the plane). The upper panels show the distortion $\rho^q_{LU}$ of the Wigner distribution, for a given transverse momentum $\vec k_\perp=k_\perp\,\vec e_y$ with $k_\perp=0.3$ GeV, induced by the proton polarization, and the lower panels show the distribution of the average quark transverse momentum. The left panels are for $u$ quarks and the right panels for $d$ quarks. These distributions have been obtained in the light-front constituent quark model.\label{fig1}}
\end{figure}

In Fig.~\ref{fig1} we show the results in impact-parameter space obtained in the light-front constituent quark model\cite{Boffi:2007yc,Pasquini:2007iz,Pasquini:2008ax,Boffi:2009sh,Pasquini:2010af,Pasquini:2011tk} (LFCQM). The upper panels show the distortions $\rho^q_{LU}$ of the Wigner distribution for $u$ (left panels) and $d$ (right panels) quarks for a given transverse momentum $\vec k_\perp=k_\perp\,\vec e_y$ with $k_\perp=0.3$ GeV. In particular, the dipole structure indicates that $u$ quarks (resp. $d$ quarks) with OAM parallel (resp. antiparallel) to the proton polarization are favored. This appears more clearly in the lower panels of Fig.~\ref{fig1} showing the impact parameter distribution of the average quark transverse momentum in a longitudinally polarized nucleon
\begin{equation}
 \langle\vec k_\perp\rangle^q(\vec b_\perp)=\int\ud x\,\ud^2k_\perp\,\vec k_\perp\,\rho^{[\gamma^+]q}_{\Lambda\Lambda}(x,\vec k_\perp,\vec b_\perp,n).
\end{equation}
Interestingly, we observe that the $d$-quark OAM changes sign around $0.25$ fm away from the transverse center of momentum. The same qualitative picture has been obtained in the light-front verion of the chiral quark-soliton model\cite{Lorce:2006nq,Lorce:2007as,Lorce:2007fa} (LF$\chi$QSM).

\section{Quark orbital angular momentum}

The Wigner distributions are rather intuitive objects as they correspond to phase-space distributions in a semiclassical picture. In particular, any matrix element of a quark operator can be rewritten as a phase-space integral of the corresponding classical quantity weighted with the Wigner distribution. It is therefore natural to define the quark OAM as follows\cite{Lorce:2011kd}
\begin{equation}\label{OAMWigner}
l_z^q=\int\ud x\,\ud^2k_\perp\,\ud^2b_\perp\left(\vec b_\perp\times\vec k_\perp\right)_z\,\rho^{[\gamma^+]q}_{\Lambda\Lambda}(x,\vec k_\perp,\vec b_\perp,n).
\end{equation}
Since the Wigner distribution involves in its definition a gauge link, it inherits a path dependence. The simplest choice is a straight gauge link. In this case, Eq.~\eqref{OAMWigner} gives the kinetic OAM $L^q_z=l^{q,\text{straight}}_z$ associated with the quark OAM operator appearing in the Ji decomposition\cite{Ji:1996ek,Ji:2012sj} $-\frac{i}{2}\int\ud^3r\,\overline\psi^q\gamma^+\left(\vec r\times\!\stackrel{\leftrightarrow}{D}_r\right)_z\psi^q$, where $D_\mu=\partial_\mu-igA_\mu$ is the usual covariant derivative. According to Ji's sum rule\cite{Ji:1996ek}, this kinetic quark OAM can be extracted from the following combination of twist-2 GPDs
\begin{equation}\label{ji-sumrule}
L^q_z=\frac{1}{2}\int^1_{-1}\ud x\left\{x\left[H^q(x,0,0)+E^q(x,0,0)\right]-\tilde H^q(x,0,0)\right\}.
\end{equation}

In order to connect the Wigner distributions to the TMDs, it is more natural to consider instead a staple-like gauge link consisting of two longitudinal straight lines connected at $x^-=\pm\infty$ by a transverse straight line\cite{Meissner:2009ww,Lorce:2011dv}. In this case, Eq.~\eqref{OAMWigner} gives the canonical OAM $\ell_z=l^{q,\text{staple}}_z$ associated with the quark OAM operator appearing in the Jaffe-Manohar decomposition\cite{Jaffe:1989jz,Lorce:2011ni,Hatta:2011ku} in the $A^+=0$ gauge $-\frac{i}{2}\int\ud^3r\,\overline\psi^q\gamma^+\left(\vec r\times\!\stackrel{\leftrightarrow}{\partial}_r\right)_z\psi^q$. Recently, it has been suggested, on the basis of some quark-model calculations, that the TMD $h_{1T}^\perp$ may also be related to the quark OAM\cite{Lorce':2011kn}
\begin{equation}
\mathcal L_z^q=-\int\ud x\,\ud^2k_\perp\,\frac{k_\perp^2}{2M^2}\,h_{1T}^{\perp q}(x,k^2_\perp).
\end{equation}
However, no rigorous expression for the OAM in terms of the TMDs is known so far. For a more detailed discussion on the different decompositions and the corresponding OAM, see Ref.~\refcite{Lorce:2012rr}.
 
\begin{table}[h!]
\tbl{Comparison between the Ji ($L^q_z$), Jaffe-Manohar ($\ell^q_z$) and TMD ($\mathcal L^q_z$) OAM in the LFCQM and the LF$\chi$QSM for $u$-, $d$- and total ($u+d$) quark contributions.}
{\begin{tabular}{@{\quad}c@{\quad}|@{\quad}c@{\quad}c@{\quad}c@{\quad}|@{\quad}c@{\quad}c@{\quad}c@{\quad}}\hline
Model&\multicolumn{3}{c@{\quad}|@{\quad}}{LFCQM}&\multicolumn{3}{c@{\quad}}{LF$\chi$QSM}\\
$q$&$u$&$d$&Total&$u$&$d$&Total\\
\hline
$L^q_z$&$0.071$&$~~0.055$&$0.126$&$-0.008$&$~~0.077$&$0.069$\\
$\ell^q_z$&$0.131$&$-0.005$&$0.126$&$~~0.073$&$-0.004$&$0.069$\\
$\mathcal L^q_z$&$0.169$&$-0.042$&$0.126$&$~~0.093$&$-0.023$&$0.069$\\
\hline
\end{tabular}\label{OAMtable}}
\end{table}

In Table \ref{OAMtable}, we present the results from the LFCQM and LF$\chi$QSM restricted to the three-quark sector\cite{Lorce:2011kd}. As expected in a pure quark model, all the definitions give the same value for the total quark OAM, with nearly twice more net quark OAM in the LFCQM than in the LF$\chi$QSM. The difference between the various definitions appears in the separate quark-flavour contributions. Note in particular that unlike the LFCQM, the LF$\chi$QSM predicts a negative sign for the $u$-quark OAM in agreement with lattice calculations\cite{Hagler:2007xi}. It is surprising that $\ell^q_z\neq L^q_z$ since it is generally believed that the Jaffe-Manohar and Ji's OAM should coincide in absence of gauge degrees of freedom. Note that a similar observation has also been made in the instant-form version of the $\chi$QSM\cite{Wakamatsu:2005vk,Wakamatsu:2009gx}. 

Recently, the twist-3 distributions attracted more attention in relation with the quark OAM. While Ji's relation is valid in the target rest frame for all three components of the angular momentum, its derivation is more natural for the transverse components in a leading-twist approach\cite{Burkardt:2002hr,Ji:2012sj}. The reason is that the OAM requires the correlation between the parton momentum and position. The GPDs encode the correlation between the longitudinal momentum and the transverse position\cite{Burkardt:2000za,Burkardt:2002hr}, and are naturally related to the transverse angular momentum. The correlation involving the transverse momentum is encoded in higher twists\cite{Penttinen:2000dg,Kiptily:2002nx,Ji:2012ba,Hatta:2012cs}. Using the parametrization from Ref.~\refcite{Meissner:2009ww} for the twist-3 GPDs, we find
\begin{equation}\label{AMSR}
-\int\ud x\,x\,\tilde E_{2T}(x,0,0)=L^q_z+2S^q_z.
\end{equation}
Combined with Ji's sum rule, it is equivalent to the Penttinen-Polyakov-Schuvaev-Strikman (PPSS) sum rule\cite{Penttinen:2000dg} obtained using another parametrization\cite{Penttinen:2000dg,Kiptily:2002nx} for the twist-3 GPDs
\begin{equation}\label{PPSS}
\int\ud x\,x\,G_2=-L^q_z.
\end{equation}
The two parametrizations are related as follows
\begin{equation}\label{reparametrization}
\begin{aligned}
H_{2T}&=-2\xi\,G_4,
&E_{2T}&=2(G_3+\xi\,G_4),\\
\tilde H_{2T}&=G_1/2,
&\tilde E_{2T}&=-(G_2+H+E)+2(G_4+\xi\,G_3),
\end{aligned}
\end{equation}
with the pure twist-3 distributions satisfying
\begin{equation}\label{momentcond}
\int\ud x\,G_i=0,\qquad\int\ud x\,x\,G_{3,4}=0.
\end{equation}
The advantage of Eq.~\eqref{AMSR} is that it involves only $\tilde E_{2T}$, which describes the vector distribution of quarks inside a longitudinally polarized target and is therefore naturally related to the $z$-component of angular momentum. Note also that the genuine angular momentum sum rule (in the sense that we can measure all terms) is actually
\begin{equation}
\int\ud x\left[x\left(H+E+2\tilde E_{2T}\right)+\tilde H\right]=0.
\end{equation}

\section{Conclusion}

In summary, we discussed the generalized transverse-momentum dependent parton distributions which gather in a single object all the information contained in the transverse-momentum dependent parton distributions and the generalized parton distributions. We interpret these mother distributions in the impact-parameter space as quark phase-space or Wigner distributions. This allowed us to write down an intuitive expression for the quark orbital angular momentum at the twist-2 level. Since, so far, it is not known how to extract these Wigner distribution from experiment, we relied on two successful relativistic light-front quark models to explore their rich phenomenology. In particular, we extracted the quark orbital angular momentum defined in terms of the Wigner distributions and compared it with alternative definitions. As expected in pure quark models, the different definitions agree on the total orbital angular momentum.

\section*{Acknowledgments}

C. Lorc\'e is thankful to INFN and the Department of Physics of the University of Pavia for their hospitality. This work was supported in part by the Research Infrastructure Integrating Activity ``Study of Strongly Interacting Matter'' (acronym HadronPhysic3, Grant Agreement n. 283286) under the Seventh Framework Programme of the European Community, by the Italian MIUR through the PRIN 2008EKLACK ``Structure of the nucleon: transverse momentum, transverse spin and orbital angular momentum'', and by the P2I (``Physique des deux Infinis'') network.


\end{document}